\documentclass{article}

\usepackage{arxiv}

\usepackage[utf8]{inputenc} 
\usepackage[T1]{fontenc}    
\usepackage{amsmath}
\usepackage{hyperref}       
\usepackage{url}            
\usepackage{booktabs}       
\usepackage{amsfonts}       
\usepackage{nicefrac}       
\usepackage{microtype}      
\usepackage{breqn}
\usepackage{graphicx}
\graphicspath{ {./images/} }
\usepackage{xcolor}
\usepackage{cite}
\usepackage{ragged2e}

\title{Traversable Casimir Wormholes in D Dimensions}

\author{
 P. H. F. Oliveira \\
  Departamento de Física, Universidade Federal do Ceará\\
  Campus do Pici, 60455-760, Fortaleza, Ceará, Brazil \\
  \href{mailto:pedrooliveira@fisica.ufc.br}{pedrooliveira@fisica.ufc.br} \\
  \And
 G. Alencar \\
 Departamento de Física, Universidade Federal do Ceará\\
 Campus do Pici, 60455-760, Fortaleza, Ceará, Brazil \\
 and
 \\
 International Institute of Physics - Federal University of Rio Grande do Norte, \\
 Campus Universitário, Lagoa Nova, Natal, RN 59078-970, Brazil\\
 \href{mailto:geova@fisica.ufc.br}{geova@fisica.ufc.br} \\
    \And
 I. C. Jardim \\
  Departamento de Física, Universidade Regional do Carirí\\
  Campus Crajubar, 63040-000, Juazeiro do Norte, Ceará, Brazil \\
  \href{mailto:ivan.jardim@urca.br}{ivan.jardim@urca.br} \\
  \And
  R. R. Landim \\
  Departamento de Física, Universidade Federal do Ceará\\
  Campus do Pici, 60455-760, Fortaleza, Ceará, Brazil \\
  \href{mailto:renan@fisica.ufc.br}{renan@fisica.ufc.br} \\
}

\begin{document}
\maketitle
\begin{abstract}
Wormholes (WH) require negative energy, and therefore an exotic matter source. Since Casimir energy is negative, it has been speculated as a good candidate to source that objects a long time ago. However only very recently a full solution for $D=4$ has been found by Garattini \cite{Garattini:2019ivd}, thus the Casimir energy can be a source of traversable WHs. Soon later Alencar \textit{et al} \cite{Alencar:2021ejd} have shown, that this is not true in $D=3$. In this paper, we show that Casimir energy can be a source of the Morris-Thorne WH for all spacetime with $D>3$. Finally, we add the cosmological constant and find that for $D=3$ Casimir WHs are possible, however, the space must always being AdS. For $D>3$, we show that the cosmological constant invert the signal with increasing throat size.
\end{abstract}

\newpage

\section{Introduction}

General Relativity, proposed by Albert Einstein in 1915, seeks to describe the gravitational interaction as a geometric effect of spacetime in response to a source of matter \cite{pad,misner1973gravitation}. Over the years it has proved to be very assertive, ensuring predictability of various phenomena \cite{Dyson:1920cwa,Ishihara:2016vdc,Carvalho:2021jlp,Abbott:2016blz}. Even in 1916, Karl Schwarzschild found the first non-trivial analytical solution for Einstein's equation, which describes the gravitational influence of a localized spherically symmetric matter \cite{Schwarzschild:1916uq}. Moreover, Schwarzschild's solution exhibits a removable singularity at a regular point, called event horizon, giving rise to the black hole idea, which was recently confirmed and photographed \cite{Akiyama:2019cqa,Akiyama:2021tfw}. By studying the interior of the event horizon, an analytic extension of Schwarzschild's coordinates reveals a linking with another asymptotically flat region beyond the fundamental singularity, $r=0$ \cite{rindler}. These structures were named  Einstein-Rosen Bridge, and connect these two separated regions in a single point, making the passage unfeasible \cite{Einstein:1935tc,hawkingellis}.  

Wormholes are solutions to Einstein's Equations and represent topological structures with a throat connecting two asymptotically flat regions of spacetime \cite{visser}.  Only in 1988, Morris and Thorne obtained the first traversable wormhole solution \cite{Morris:1988cz}. In order to rule the traversability of a Wormhole, two geometric conditions are imposed, the flare-out condition \cite{Hochberg:1998ha}, which is related to the minimization of the throat, and the finite redshift condition. In general, they do not satisfy the energy conditions of the General Relativity, being necessary some type of exotic matter as source \cite{Bronnikov:2014bda, Kim:2013tsa}. However, from the embedding procedure in cylindrical coordinates, Kim \cite{Kim:2013tsa} showed that a negative energy density satisfies the strong flare-out condition for the Morris-Thorne wormhole type, although it is not necessary for the weak condition \cite{Eiroa:2009nn}. The understanding of such solutions has become the topic of intense research in recent years\cite{Gao:2016bin}, including the discovery of humanly traversable wormholes in Ref. \cite{Maldacena:2020sxe}.

The Casimir effect involves negative energies of quantum fields\cite{Moste}. The effects of curved spacetime in this vacuum energy is presently an object of discussion in the literature \cite{Sorge:2005ed,Lima:2019pbo,Lima:2020igm}. This issue has drawn attention since the proposal of observational investigation, in the Archimedes vacuum weight experiment\cite{Avino:2019fdq}. Due to the negative energy,  it was sought as a candidate to produce a traversable wormhole a long time ago by Morris and Thorn and sometime later by Visser  \cite{Morris:1988tu,visser}. However only very recently a full solution has been found by Garattini\cite{Garattini:2019ivd}. The author found an appropriate redshift function and proved that in four-dimensional spacetime, the Casimir energy, solely, can be a source of traversable wormholes. After this seminal work, many consequences has been studied \cite{Carvalho:2021ajy,Javed:2020mjb,Garattini:2020kqb,Tripathy:2020ehi,Jusufi:2020rpw,Alencar:2021ejd}. In this direction  Alencar {\it et al}   have shown that, differently of the case with $D=4$,   in $D=3$ Casimir energy, solely,  can not be a source of wormholes and additional sources must be added \cite{Alencar:2021ejd}. This raises the question if Casimir energy can be a source in the case of arbitrary dimensions. This becomes particularly important in the case $D=5$, where Casimir energy is a necessary ingredient for the aforementioned humanly traversable wormholes\cite{Maldacena:2020sxe}.

In this work, we propose to study the influence of dimensionality of spacetime on the Casimir Wormhole, in other to determine a general condition of their traversability. We will do this analysis solving Einstein's equation in $D$ dimensions static and spherically symmetric spacetime and considering the Casimir energy and pressure as sources. Furthermore, we will discuss the implications of considering the cosmological constant as an additional source of the Casimir wormhole.

\section{Traversable Casimir Wormholes in D Dimensions}

In this section, we will study the existence of Traversable Casimir Wormholes in arbitrary dimensions. In four dimensions Garattini shows that just the standard negative Casimir energy density and the radial pressure are able to form a Morris-Thorne Wormhole \cite{Garattini:2019ivd}. In another hand, in a recent paper, the authors show that in three dimensions the same is not true \cite{Alencar:2021ejd}, thus we will investigate now, the influence of dimensionality of spacetime on the Casimir wormhole, in order to generalize both results.

The study of wormholes in the context of extra dimensions has been extensively studied in Ref's. \cite{Kar:1995ss,DeBenedictis:2002wd,Cataldo:2002jw}. Since the Morris-Thorne wormholes is static and spherically symmetric, we take the general metric in $D$ dimensions, following \cite{Morris:1988cz,Gogberashvili:1998iu,Das:2001md,Jardim:2011gg}

\begin{equation}\label{lineelement}
	ds^2 = -e^{2\Phi(r)}dt^2 + \frac{1}{1 - \frac{b(r)}{r}}dr^2 + r^2d\Omega^2_{(D - 2)},
\end{equation}
where $d\Omega^2_{(D - 2)}$ is the line element of a unit sphere of $D - 2$ dimensions \cite{Das:2001md}; $\Phi(r)$ and $b(r)$ are the redshift and shape functions, respectively, both are arbitrary functions of $r\in[r_0,+\infty)$, with $r_0$ being the radius from the wormholes throat. The independent components of Einstein's equation in orthonormal base, described by line element (\ref{lineelement}), are

\begin{subequations}
	\begin{eqnarray}
		\kappa_D\rho_D(r) &=& \frac{D - 2}{2r^3}\left[(D - 4)b + rb'\right],\label{Einstein1}\\
		\kappa_D\tau_D(r) &=&-\frac{D - 2}{2r^3}\left[(D-3)b+2r(b-r)\Phi'\right],\label{Einstein2}\\
		\kappa_Dp_D(r)&=&\left(1-\frac{b}{r}\right)\left[\Phi''+\Phi'\left(\Phi'+\frac{D-3}{r}\right)\right] - \frac{3(D-4)(D-3)}{2r^2}-\frac{rb'}{2r^2}\left[\Phi'+\frac{D-3}{r}\right]\nonumber\\&&+\frac{b}{2r^2}\left[\Phi'-\frac{(D-5)(D-3)}{r}\right],\label{Einstein3}
	\end{eqnarray}
\end{subequations}
where $(')$ means the derivative with respect to $r$; $\kappa_D$ is the gravitational coupling constant in $D$ dimensions; $\rho_D(r)$ is the surface energy density, $\tau_D(r)$ and $p_D(r)$ are the radial and transverse pressure, respectively.

According to Eq. (\ref{Einstein1}), the flare-out condition, $b'r - b < 0$, is satisfied for any dimensions if $\rho_D(r) < 0$. Since the Casimir energy density of a massless field is a typical system with negative energy, it should be possible to build a wormhole, a priori, in any dimension. For arbitrary dimensions, the Casimir energy is given by (in natural units) \cite{Alnes:2006pa}
\begin{equation}\label{energycasimir}
	\rho_D(r)=-\frac{(D - 2)\Gamma(D/2)\zeta_R(D)}{(4\pi)^{D/2}}\frac{1}{r^D}=-\frac{\lambda_D}{r^D},
\end{equation}
where $\zeta_R(z)$ is the Riemann zeta function. The $D - 2$ factor appears due to the $D - 2$ degrees of freedom of the electromagnetic field. Taking $D = 4$, we recover the original result \cite{Casimir:1948dh}. Furthermore, the radial pressure is given by the Equation of State (EoS)
\begin{equation}\label{EoS}
	\tau_D(r) = (D - 1)\rho_D(r) = \omega_D\rho_D(r).
\end{equation}

In order to obtain the shape function, $b(r)$, we use the Casimir energy density (\ref{energycasimir}) to write the Eq. (\ref{Einstein1}) in the form
\begin{equation}
	\left[r^{D - 4}b(r)\right]' = - \frac{r_1^{D - 2}}{r^2},
\end{equation}
where, generalizing Garattini's notation \cite{Garattini:2019ivd}, we have defined
\begin{equation}
	r_1^{D - 2} \equiv \frac{2\kappa_D\lambda_D}{D - 2}.
\end{equation}

Fixing the throat of wormhole at $r = r_0$, i.e., $b(r_0) = r_0$, the shape function is given by
\begin{equation}\label{shape}
	b(r) = \frac{r_1^{D - 2}}{r^{D - 3}} + \frac{r_0^{D - 3}}{r^{D-4}}\left[1 - \frac{r_1^{D-2}}{r_0^{D - 2}}\right],
\end{equation}
which is a consistent generalization for arbitrary dimensions of the shape functions obtained in literature for $D = 3$ \cite{Alencar:2021ejd} and $D = 4$ \cite{Garattini:2019ivd}. Furthermore $\lim\limits_{r\rightarrow\infty}b(r) = 0$, ensuring that the solution asymptotically returns to Minkowski's.

To compute the redshift function we use the EoS (\ref{EoS}) and the shape function (\ref{shape}) to write the Eq. (\ref{Einstein2}) in the compact form
\begin{equation}\label{edophi}
	2r_0r^{D - 2}(b - r)\Phi' = \left[\omega_D - (D - 3)\right]r_0r_1^{D - 2} - (D - 3)\left[r_0^{D - 2} - r_1^{D - 2}\right]r,
\end{equation}
despite the above equation can not be solved for arbitrary dimensions, we can analyze the divergence of $\Phi(r)$. First, let us turn our attention to the term $b - r$ in left hand side of Eq. (\ref{edophi}), it can be factored as
\begin{equation}\label{b-r}
	b - r = -(r - r_0)\frac{P_{D - 3}(r)}{r^{D - 3}},
\end{equation}
where did we have use the identity
\begin{equation}
	x^n - a^n = (x - a)\sum_{i=0}^{n - 1}x^{n -1 - i}a^i,
\end{equation}
for $n > 0$, and defined the rank $D - 3$ polynomial, $P_{D - 3}(r)$, as
\begin{equation}\label{poly}
	P_D(r) = \left\{\begin{array}{l}
		r_1^{D - 2}r_0^{-1} \ \textnormal{, if $D = 3$}\\
		r_1^{D-2}r_0^{-1} + \sum_{i=0}^{D - 4}r^{D - 3 - i}r_0^i \ \textnormal{, if $D > 3$}.
	\end{array}\right.
\end{equation}

Since $P_{D - 3}(r)$ has only positive terms, it has no positive real roots \cite{conway}, therefore not being a physical divergence term. Thus the vanishes of $b - r$ term at the throat occurs only due the $r - r_0$ factor in right hand side of Eq. (\ref{b-r}). Then, the Eq. (\ref{edophi}) take the form
\begin{equation}\label{phi'}
	\Phi' = \frac{\left[\omega_D - (D - 3)\right]r_0r_1^{D - 2} - (D - 3)\left[r_0^{D - 2} - r_1^{D - 2}\right]r}{2r_0r(r - r_0)P_{D - 3}(r)}.
\end{equation}

To avoid the divergence at wormhole throat, $r = r_0$, the first order polynomial in the numerator must to vanishes at this point. Thus, due the freedom in fixing $r_0$, we get
\begin{equation}\label{omegaD}
	\omega_D = (D - 3)\frac{r_0^{D - 2}}{r_1^{D - 2}},
\end{equation}
for $D = 4$ the results of Garattini are retrieved \cite{Garattini:2019ivd}. So, note that only for $D = 3$, is not possible to use the fixing (\ref{omegaD}) to remove the divergence in Eq. (\ref{phi'}), keeping the standard EoS (\ref{EoS}). Therefore, for $D > 3$ is always possible to use the Casimir energy as the source of a wormhole and the redshift function is given by solution of the following ODE
\begin{equation}\label{redshift}
	\Phi' = \frac{(D - 3)\left[r_0^{D - 2} - r_1^{D - 2}\right]}{2r_0rP_{D - 3}(r)}.
\end{equation}

Meanwhile, we must show that the solution at the asymptotic limit returns Minkowski spacetime, given the definition of a wormhole. For this, we must compute
\begin{equation}\label{fixingphi0}
	\lim\limits_{r\rightarrow\infty}\Phi(r) = \Phi(r_0) + \frac{(D - 3)\left[r_0^{D - 2} - r_1^{D - 2}\right]}{2r_0}\lim\limits_{r\rightarrow\infty}\int_{r_0}^{r}\frac{dr'}{r'P_{D - 3}(r')},
\end{equation}
where $\Phi(r_0)$ is an integration constant. In fact, since the roots of $P_{D - 3}(r)$ are not positive real, then the integrand has no physical divergences since $r = 0$ is not comprised in the domain of $\Phi(r)$, thus, for $D > 3$ the integral converge. So, we can fix $\Phi(r_0)$ to obtain the Minkowski spacetime as an asymptotic limit.

Therefore, for every spacetime of dimension $D>3$, it is possible to construct a traversable Casimir wormhole of the Morris-Thorne type, with a shape function given by Eq. (\ref{shape}) and a redshift function given by Eq. (\ref{redshift}).

\section{Casimir Wormholes with cosmological constant}

\begin{figure}[!h]
	\centering
	\includegraphics[width=0.6\linewidth]{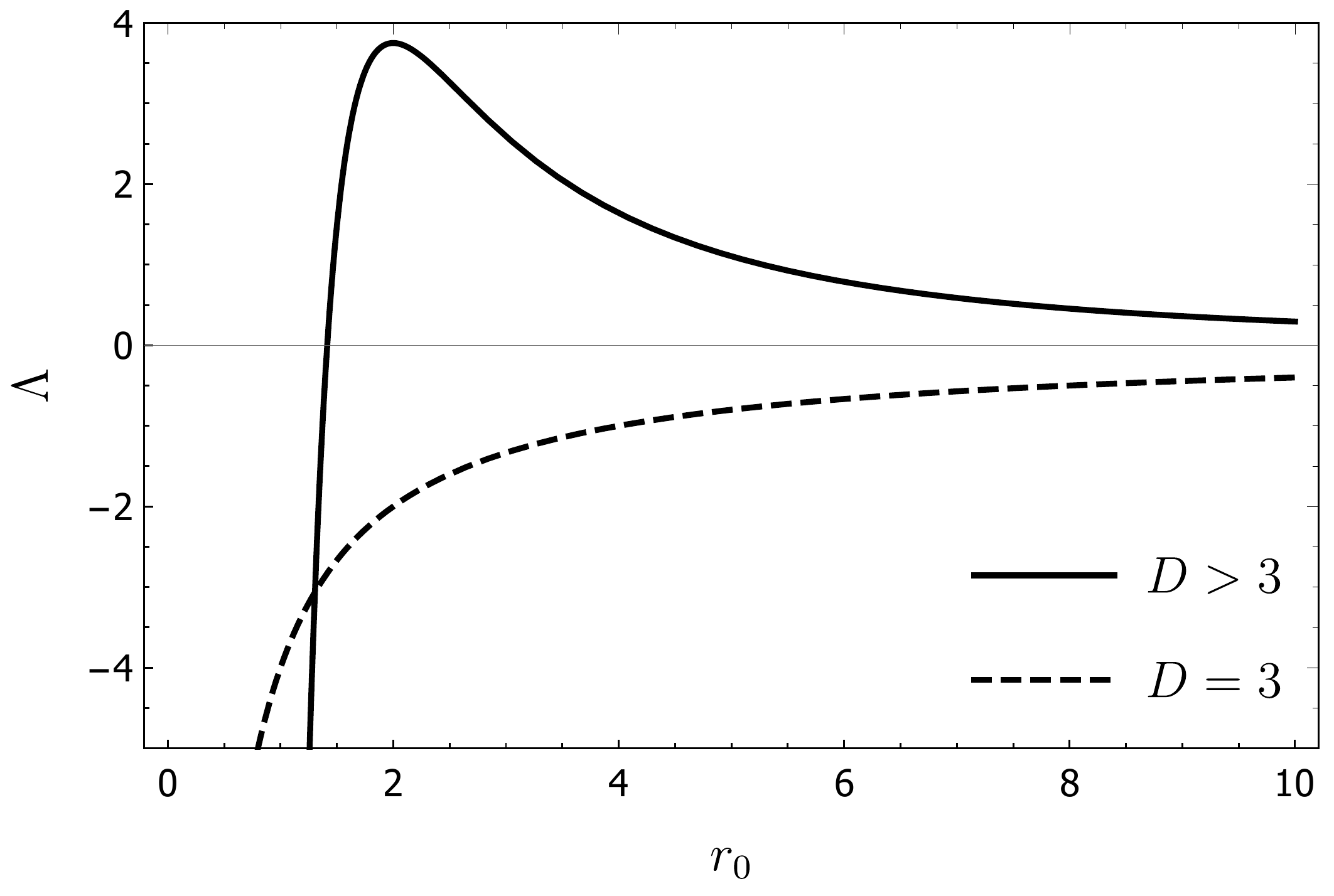}
	\caption{Relation among cosmological constant, $\Lambda$, and throat size of Casimir Wormhole, $r_0$, for $D = 3$ (dashed line) and $D > 3$ (solid line).}
	\label{fig:graph}
\end{figure}

A possibility to get around the inexistence of a Casimir Wormhole for $D = 3$ is proposed by Alencar \textit{et al} \cite{Alencar:2021ejd}, adding the cosmological constant in the moment-energy tensor. Einstein's equation with the cosmological constant is modified by,
\begin{equation}
	R^N_M - \frac{1}{2}R\delta^N_M + \Lambda\delta^N_M = \kappa_DT^N_M,
\end{equation}
where the capital latin indices vary under the $D$ dimensions in the line element given by Eq. (\ref{lineelement}). Thus, the following shape function is obtained
\begin{equation}\label{shapecosmol}
	b_\Lambda(r) = b(r) + \frac{2\Lambda\left(r^{D - 1} - r_0^{D - 1}\right)}{(D-2)(D-1)r^{D - 4}},
\end{equation}
where $b(r)$ is given by Eq. (\ref{shape}). On the other hand, to circumvent the divergence in the redshift function in the wormhole's throat we must fix
\begin{equation}
	\omega_D = (D - 3)\frac{r_0^{D - 2}}{r_1^{D - 2}} - \frac{2\Lambda}{(D - 2)}\frac{r_0^D}{r_1^{D - 2}},
\end{equation}
thus, if we fix $\omega_D = D - 1$ which is the condition of the EoS for the energy of Casimir (\ref{EoS}), then we get a relation among the cosmological constant and the radius of the wormhole throat,
\begin{equation}
	\Lambda(r_0) = \frac{\alpha_D}{r_0^2} - \frac{\beta_D}{r_0^{D}},
\end{equation}
where $\alpha_D = (D - 3)(D - 2)/2$, and $\beta_D = 8\pi(D - 1)G_D\lambda_D$, and $G_D$ is the gravitational constant in $D$ dimensions, given by (in natural units) $G_{(D = d + 1)} = 2\pi^{1 - d/2}\Gamma(d/2)l_P^{d-1}$ \cite{Mann:2011rh}, and $l_P$ is the Planck length, and $\lambda_D$ is the Casimir energy constant defined by Eq. (\ref{energycasimir}). Fig. (\ref{fig:graph}) shows the behavior of the cosmological constant value versus the throat size. 

Note that for $D = 3$ the cosmological constant is necessarily negative, so the $(2+1)$-dimensional spacetime is of the Anti-deSitter (AdS) type. On the other hand, for $D > 3$ the cosmological constant has a real Planckian root, characterizing an inversion in the sign. Thus, for wormholes with throats smaller than $r_0 = (\beta_D/\alpha_D)^{1/(D-2)}$ the cosmological constant is negative, while for larger values it must have $\Lambda > 0$, with a maximum at $r_0 = (D\beta_D/2\alpha_D)^{1/(D - 2)}$ and tending to zero for $r_0 \rightarrow \infty$. 

If we choose a small positive cosmological constant, i.e. de Sitter spacetime, in $D = 4$ the cosmological constant value is $\Lambda\approx10^{-52} m^{-2}$ \cite{Carmeli:2000cf}, then we have two possibilities, or the wormhole is planckian ($r_0\approx1.0186 l_P$) or macroscopic ($r_0\approx10^{26} m$) size. Thus, for the scale of energy considered of source, the planckian size is the unique physical solution reasonable, while for $\Lambda < 0$ all the wormholes is subplanckian. On the other hand, in a context of small (negative) cosmological constant at $D = 3$ it is necessarily related to a wormhole of macroscopic size. Also, the limit of $|\Lambda|\rightarrow\infty$ has the same behavior independently of dimension, we get $r_0\rightarrow0$.

Although for the case of cosmological constant it is not possible to recover the minkowskian limit, we have deSitter or Anti-deSitter type spacetimes, the redshift function is determined by solving the following ODE
\begin{equation}
	-2rP_{D - 3}^{(\Lambda)}\Phi' = \frac{2\Lambda}{(D - 2)}\sum_{j = 0}^{D - 1}r^{D -1 - j}r_0^j + (D - 3)r_1^{D - 2}r_0^{-1} - (D - 3)r_0^{D - 3} - \frac{2(D - 3)\Lambda r}{(D - 2)(D - 1)}\sum_{j = 0}^{D - 2}r^{D - 2 - j}r_0^j,
\end{equation}
where we redefine the $D-3$ polynomial of the Eq. (\ref{poly}) to
\begin{equation}
	P^{(\Lambda)}_{D - 3}(r) = P_{D - 3}(r) - \frac{2\Lambda}{(D - 2)(D - 1)}\sum_{j = 0}^{D - 2}r^{D -2 - j}r_0^j.
\end{equation}

Thus, in cases of negative cosmological constant, i.e., wormholes of Planckian size, $\Phi(r)$ does not have singularities in the range $[r_0,\infty)$. However, in the context of $\Lambda > 0$ the redshift function presents a divergence characterized by de Sitter horizon.

\section{Conclusion}

In this article, we study the formation of static, spherically symmetric and traversable wormholes for spacetimes of arbitrary dimension, generalizing the recent works \cite{Alencar:2021ejd} and \cite{Garattini:2019ivd}. For this, we analyzed Einstein's equation in $D$ dimensions static and spherically symmetric spacetime and considering the Casimir energy and pressure as sources.  We found, similarly to Refs.  \cite{Garattini:2019ivd,Alencar:2021ejd}, that the solution depends only on the redshift and shape functions. To guarantee the flare-out and the finite redshift condition, we analyzed the radius of the throat and the dimensions that permit a fixing that removes the roots in the redshift equation. 

We obtained an analytical expression for the shape function for D arbitrary (\ref{shape}). It  reproduces the expressions found for $D = 3$ \cite{Alencar:2021ejd} and $D = 4$ \cite{Garattini:2019ivd}, in addition to being asymptotically null. In fact, we show that the boundary and flare-out conditions can be satisfied for any $D$. Therefore these conditions do not impose any restrictions for Casimir wormholes in $D$ dimensions.

On the other hand, in order to be traversable, we must find a wormhole solution that avoids an event horizon. This is exactly where a restriction appears. In Ref \cite{Alencar:2021ejd} the authors show that, differently of the $D=4$ case\cite{Garattini:2019ivd}, a divergence in $r = r_0$ in the redshift parameter is inescapable if the Casimir energy is the only source. To find the origin of this singularity,  we obtained a general condition to be satisfied by the state parameter $\omega_D$ to avoid it. It is given by
\begin{equation}\label{omegaDConc}
	\omega_D = (D - 3)\frac{r_0^{D - 2}}{r_1^{D - 2}},
\end{equation}
and this expression reveals that for $D=3$ we must have $\omega_3 = 0$, eliminating the Casimir source. This explains the reason why for three-dimensional spacetimes it is not possible to form a wormhole, a result formerly presented in Ref. \cite{Alencar:2021ejd}. In parallel, we have shown that, although it was not possible to obtain an analytic expression for $\Phi(r)$, the integrand does not have divergences across the domain when $D > 3$. This guaranteed that $\Phi(r) $ is finite everywhere and that we have no horizon. We also show that, in the asymptotic limit, the integral converges allowing a fixation on $\Phi(r_0)$ (\ref{fixingphi0}) and guaranteeing that the solution is asymptotically flat. 

Based on the above results, we conclude that the Casimir energy can be used as a Morris-Thorne wormhole source for every spacetime of dimension $D>3$. We should point that this must have consequences for the physics of extra dimensions, particularly related to the newly discover humanly traversable wormhole of Ref. \cite{Maldacena:2020sxe}.

Finally, we discuss the consequences of considering the existence of a non-zero cosmological constant. In this scenario we got a shape function (\ref{shapecosmol}) that satisfies all the conditions imposed. We show that to circumvent the throat divergence we must fix a relation among the value of the cosmological constant and the radius of the throat by
\begin{equation}\label{result}
	\Lambda(r_0) = \frac{(D - 3)(D - 2)}{2r_0^2} - \frac{8\pi(D - 1)G_D\lambda_D}{r_0^{D}}.
\end{equation}
From the above expression, we see that in $D=3$ a Casimir wormhole is possible, but the spacetime must by AdS. This is very similar to the BTZ black hole,  where a negative cosmological constant must be added in order provide a non-trivial solution \cite{Banados:1992wn}. For $D>3$ we find that both possibilities, de Sitter and anti de Sitter,  are allowed. The above expression also shows that the signal of the cosmological constant is controlled by the radius of the throat. The case $\Lambda=0$ for $D = 4$ also gives the value of $r_0$ previously found in Ref.\cite{Garattini:2019ivd}. Finally, Eq. (\ref{result}) shows that, for $D > 3$, in the context of a de Sitter space with a small cosmological constant, there are two options for $r_0$, but only planckian solution is physically reasonable.

\section*{Acknowledgement}

The authors would like to thanks the financial support provided by the Coordenação de Aperfeiçoamento de Pessoal de Nível Superior (CAPES) and by the Conselho Nacional de Desenvolvimento Científico e Tecnológico (CNPq) through Universal 426540/2018-2 and 424556/2018-9. We also acknowledge Fundação Cearense de Apoio ao Desenvolvimento Científico e Tecnológico (FUNCAP) through PRONEM PNE0112-00085.01.00/16.


\begin{thebibliography}{00}
\justifying

\bibitem{Garattini:2019ivd}
R.~Garattini,
Eur. Phys. J. C \textbf{79}, no.11, 951 (2019)
doi:10.1140/epjc/s10052-019-7468-y
[arXiv:1907.03623 [gr-qc]].

\bibitem{Alencar:2021ejd}
G.~Alencar, V.~B.~Bezerra and C.~R.~Muniz,
[arXiv:2104.13952 [gr-qc]].

\bibitem{pad} T. Padmanabhan, Gravitation, first ed., Cambridge University Press,  Cambridge, 2010.

\bibitem{misner1973gravitation} C. W. Misner, K. S. Thorne and J. A. Wheeler, Gravitation, first ed., W. H. Freeman, San Francisco, 1973.

\bibitem{Dyson:1920cwa}
F.~W.~Dyson, A.~S.~Eddington and C.~Davidson,
Phil. Trans. Roy. Soc. Lond. A \textbf{220}, 291-333 (1920)
doi:10.1098/rsta.1920.0009

\bibitem{Ishihara:2016vdc}
A.~Ishihara, Y.~Suzuki, T.~Ono, T.~Kitamura and H.~Asada,
Phys. Rev. D \textbf{94}, no.8, 084015 (2016)
doi:10.1103/PhysRevD.94.084015
[arXiv:1604.08308 [gr-qc]].

\bibitem{Carvalho:2021jlp}
I.~D.~D.~Carvalho, G.~Alencar, W.~M.~Mendes and R.~R.~Landim,
[arXiv:2103.03845 [gr-qc]].

\bibitem{Abbott:2016blz}
B.~P.~Abbott \textit{et al.} [LIGO Scientific and Virgo],
Phys. Rev. Lett. \textbf{116}, no.6, 061102 (2016)
doi:10.1103/PhysRevLett.116.061102
[arXiv:1602.03837 [gr-qc]].
%

\bibitem{Schwarzschild:1916uq}
K.~Schwarzschild,
Sitzungsber. Preuss. Akad. Wiss. Berlin (Math. Phys.) \textbf{1916}, 189-196 (1916)
[arXiv:physics/9905030 [physics]].

\bibitem{Akiyama:2019cqa}
K.~Akiyama \textit{et al.} [Event Horizon Telescope],
Astrophys. J. Lett. \textbf{875}, L1 (2019)
doi:10.3847/2041-8213/ab0ec7
[arXiv:1906.11238 [astro-ph.GA]].

\bibitem{Akiyama:2021tfw}
K.~Akiyama, J.~C.~Algaba, A.~Alberdi, W.~Alef, R.~Anantua, K.~Asada, R.~Azulay, A.~K.~Baczko, D.~Ball and M.~Balokovi\'c, \textit{et al.}
Astrophys. J. Lett. \textbf{910}, no.1, L13 (2021)
doi:10.3847/2041-8213/abe4de
[arXiv:2105.01173 [astro-ph.HE]].

\bibitem{rindler} W. Rindler, Relativity: special, general and cosmological, second ed., Oxford University Press, Oxford, 2006.

\bibitem{Einstein:1935tc}
A.~Einstein and N.~Rosen,
Phys. Rev. \textbf{48}, 73-77 (1935)
doi:10.1103/PhysRev.48.73

\bibitem{hawkingellis} S. W. Hawking and G. F. R. Ellis, The Large Scale Structure of Space-Time, first ed., Cambridge University Press, Cambridge, 1973.

\bibitem{visser} M. Visser, Lorentzian Wormholes: From Einstein to Hawking, first ed., American Institute of Physics, New York, 1996.

\bibitem{Morris:1988cz}
M.~S.~Morris and K.~S.~Thorne,
Am. J. Phys. \textbf{56}, 395-412 (1988)
doi:10.1119/1.15620

\bibitem{Hochberg:1998ha}
D.~Hochberg and M.~Visser,
Phys. Rev. D \textbf{58}, 044021 (1998)
doi:10.1103/PhysRevD.58.044021
[arXiv:gr-qc/9802046 [gr-qc]].

\bibitem{Bronnikov:2014bda}
K.~A.~Bronnikov and M.~V.~Skvortsova,
Grav. Cosmol. \textbf{20}, 171-175 (2014)
doi:10.1134/S0202289314030062
[arXiv:1404.5750 [gr-qc]].

\bibitem{Kim:2013tsa}
S.~W.~Kim,
J. Korean Phys. Soc. \textbf{63}, 1887-1891 (2013)
doi:10.3938/jkps.63.1887
[arXiv:1302.3337 [gr-qc]].

\bibitem{Eiroa:2009nn}
E.~F.~Eiroa and C.~Simeone,
Phys. Rev. D \textbf{81}, no.8, 084022 (2010)
[erratum: Phys. Rev. D \textbf{90}, no.8, 089906 (2014)]
doi:10.1103/PhysRevD.81.084022
[arXiv:0912.5496 [gr-qc]].

\bibitem{Gao:2016bin}
P.~Gao, D.~L.~Jafferis and A.~C.~Wall,
JHEP \textbf{12}, 151 (2017)
doi:10.1007/JHEP12(2017)151
[arXiv:1608.05687 [hep-th]].

\bibitem{Maldacena:2020sxe}
J.~Maldacena and A.~Milekhin,
Phys. Rev. D \textbf{103}, no.6, 066007 (2021)
doi:10.1103/PhysRevD.103.066007
[arXiv:2008.06618 [hep-th]].

\bibitem{Moste} M. Bordag, G. L. Klimchitskaya, U. Mohideen, and V. M. Mostepanenko, Advances in the Casimir Effect, first ed., Oxford Science Publications, Oxford, 2015.

\bibitem{Lima:2019pbo}
A.~P.~C.~M.~Lima, G.~Alencar, C.~R.~Muniz and R.~R.~Landim,
JCAP \textbf{07}, 011 (2019)
doi:10.1088/1475-7516/2019/07/011
[arXiv:1903.00512 [hep-th]].

\bibitem{Sorge:2005ed}
F.~Sorge,
Class. Quant. Grav. \textbf{22}, 5109-5119 (2005)
doi:10.1088/0264-9381/22/23/012

\bibitem{Lima:2020igm}
A.~P.~C.~M.~Lima, G.~Alencar and R.~R.~Landim,
JCAP \textbf{01}, 056 (2021)
doi:10.1088/1475-7516/2021/01/056
[arXiv:2007.07163 [hep-th]].

\bibitem{Avino:2019fdq}
S.~Avino, E.~Calloni, S.~Caprara, M.~De Laurentis, R.~De Rosa, T.~Di Girolamo, L.~Errico, G.~Gagliardi, M.~Grilli and V.~Mangano, \textit{et al.}
MDPI Physics \textbf{2}, no.1, 1-13 (2020)
doi:10.3390/physics2010001


\bibitem{Morris:1988tu}
M.~S.~Morris, K.~S.~Thorne and U.~Yurtsever,
Phys. Rev. Lett. \textbf{61}, 1446-1449 (1988)
doi:10.1103/PhysRevLett.61.1446

\bibitem{Carvalho:2021ajy}
I.~D.~D.~Carvalho, G.~Alencar and C.~R.~Muniz,
[arXiv:2106.11801 [gr-qc]].

\bibitem{Javed:2020mjb}
W.~Javed, A.~Hamza and A.~\"Ovg\"un,
Mod. Phys. Lett. A \textbf{35}, no.39, 2050322 (2020)
doi:10.1142/S0217732320503228
[arXiv:2101.04515 [gr-qc]].

\bibitem{Garattini:2020kqb}
R.~Garattini,
Eur. Phys. J. C \textbf{80}, no.12, 1172 (2020)
doi:10.1140/epjc/s10052-020-08728-8
[arXiv:2008.05901 [gr-qc]].

\bibitem{Tripathy:2020ehi}
S.~K.~Tripathy,
Phys. Dark Univ. \textbf{31}, 100757 (2021)
doi:10.1016/j.dark.2020.100757
[arXiv:2004.14801 [gr-qc]].

\bibitem{Jusufi:2020rpw}
K.~Jusufi, P.~Channuie and M.~Jamil,
Eur. Phys. J. C \textbf{80}, no.2, 127 (2020)
doi:10.1140/epjc/s10052-020-7690-7
[arXiv:2002.01341 [gr-qc]].

\bibitem{Kar:1995ss}
S.~Kar and D.~Sahdev,
Phys. Rev. D \textbf{53}, 722-730 (1996)
doi:10.1103/PhysRevD.53.722
[arXiv:gr-qc/9506094 [gr-qc]].

\bibitem{DeBenedictis:2002wd}
A.~DeBenedictis and A.~Das,
Nucl. Phys. B \textbf{653}, 279-304 (2003)
doi:10.1016/S0550-3213(03)00051-8
[arXiv:gr-qc/0207077 [gr-qc]].

\bibitem{Cataldo:2002jw}
M.~Cataldo, P.~Salgado and P.~Minning,
Phys. Rev. D \textbf{66}, 124008 (2002)
doi:10.1103/PhysRevD.66.124008
[arXiv:hep-th/0210142 [hep-th]].

\bibitem{Gogberashvili:1998iu}
M.~Gogberashvili,
Europhys. Lett. \textbf{49}, 396-399 (2000)
doi:10.1209/epl/i2000-00162-1
[arXiv:hep-ph/9812365 [hep-ph]].

\bibitem{Das:2001md}
A.~Das and A.~DeBenedictis,
Prog. Theor. Phys. \textbf{108}, 119-132 (2002)
doi:10.1143/PTP.108.119
[arXiv:gr-qc/0110083 [gr-qc]].

\bibitem{Jardim:2011gg}
I.~C.~Jardim, R.~R.~Landim, G.~Alencar and R.~N.~Costa Filho,
Phys. Rev. D \textbf{84}, 064019 (2011)
doi:10.1103/PhysRevD.84.064019
[arXiv:1105.4578 [gr-qc]].

\bibitem{Alnes:2006pa}
H.~Alnes, F.~Ravndal, I.~K.~Wehus and K.~Olaussen,
Phys. Rev. D \textbf{74}, 105017 (2006)
doi:10.1103/PhysRevD.74.105017
[arXiv:quant-ph/0607081 [quant-ph]].

\bibitem{Casimir:1948dh}
H.~B.~G.~Casimir,
Indag. Math. \textbf{10}, 261-263 (1948)

\bibitem{conway} J. B. Conway, Functions of One Complex Variable I, second ed., Springer Science \& Business Media, New York, 1978.

\bibitem{Mann:2011rh}
R.~B.~Mann and J.~R.~Mureika,
Phys. Lett. B \textbf{703}, 167-171 (2011)
doi:10.1016/j.physletb.2011.07.052
[arXiv:1105.5925 [hep-th]].

\bibitem{Carmeli:2000cf}
M.~Carmeli and T.~Kuzmenko,
AIP Conf. Proc. \textbf{586}, no.1, 316-318 (2001)
doi:10.1063/1.1419571
[arXiv:astro-ph/0102033 [astro-ph]].

\bibitem{Banados:1992wn}
M.~Banados, C.~Teitelboim and J.~Zanelli,
Phys. Rev. Lett. \textbf{69}, 1849-1851 (1992)
doi:10.1103/PhysRevLett.69.1849
[arXiv:hep-th/9204099 [hep-th]].

\end{thebibliography}

\end{document}